\documentclass[aps,prb,twocolumn,superscriptaddress,raggedbottom,showpacs]{revtex4}
\usepackage{graphicx,latexsym}
\usepackage{amsmath}
\usepackage{dcolumn}
\usepackage{bm}
\begin{document}
\title
{\bf Non-adiabatic transport in a quantum dot turnstile }
\author{Valeriu Moldoveanu}
\affiliation{National Institute of Materials Physics, P.O. Box MG-7,
Bucharest-Magurele, Romania}
\author{Vidar Gudmundsson}
\affiliation{Science Institute, University of Iceland, Dunhaga 3, IS-107 Reykjavik, Iceland}
\author{Andrei Manolescu}
\affiliation{National Institute of Materials Physics, P.O. Box MG-7,
Bucharest-Magurele, Romania}

\begin{abstract} 

We present a theoretical study of the electronic transport 
through a many-level quantum dot driven by time-dependent signals applied at the contacts to the leads.
If the barriers oscillate out of phase the system operates 
like a turnstile pump under a finite constant bias, as 
observed in the experiments of Kouwenhoven {\it et al.} [Phys. Rev. Lett. {\bf 67}, 1626 (1991)]. 
The time-dependent 
currents and their averages over succesive pumping periods are computed from the Keldysh formalism
for tight-binding models. 
The calculation considers a sudden application of the pumping potentials at $t=0$ which 
leads to transient features 
of the time-dependent and averaged currents during the first pumping
cycles which turn out to be important in the high-frequency regime. 
We show that in the transient regime
the efficiency of the system as a pump is rather poor because it mainly absorbs charge 
from both leads in
order to fill the levels located
below the bias window.
Under a finite bias and a low-frequency pumping signal
 the charge transferred across the system depends on the number of levels located within the bias window.
The internal charge dynamics and the role of energy sidebands are investigated. The so called satellite peaks of the averaged current 
are observed
also in the transient regime.

\end{abstract}


\maketitle

\section{Introduction}

The ability to control the transport properties of semiconductor quantum dots by 
time-dependent perturbations (e.g.\ microwave signals or optical pulses) allows the 
observation of photon-assisted tunneling, charge pumping \cite{Geerligs,Switkes,Kou} 
 and coherent Rabi oscillations. \cite{Zrenner}
Also, pump-and-probe techniques were used to estimate relaxation rates and to control 
spins in quantum dots. \cite{Tarucha}  A common point of these experiments is that 
the time-dependent driving potential
is applied {\it on} the system itself, i.e.\ on a central metalic gate defining the quantum dot.

Some time ago Kouwenhoven {\it et al.} \cite{TSP} proposed a different setup, in which a quantum 
dot is coupled to 
source and drain reservoirs by oscillating tunneling barriers. Technically this is achieved by 
applying radio-frequency 
signals to the metalic gates that control the opening of the quantum dot to its 
surroundings. The two barriers at the contacts are varied in such a way that the system undergoes a cyclic
transformation and, under a constant bias applied on the leads, an integer number of electrons is transmitted
during one cycle. Therefore the system operates as a turnstile pump. An important feature of 
the turnstile configuration is that the pumped current has a definite direction due to the finite bias.
We remind here that originally the concept of parametric charge pumping was introduced in the context 
of a net current 
generation in an {\it unbiased} system. \cite{Brower} As noted in the literature, a symmetry breaking is necessary in order
to get a nonvanishing current without a bias. In spite of the fact that the turnstile operation was 
experimentally observed some time ago, it attracted little attention in the theoretical literature 
(see Refs.\ \onlinecite{Aharony,SL,Braginsky,Arrachea} below). The
purpose of this work is to explore the transport properties of turnstile quantum dots submitted to
time-dependent signals of arbitrary amplitude and frequency.   

At theoretical level the quantum pumping was discussed basically within two frameworks: the adiabatic or 
Floquet scattering theory \cite{Brower,Levinson,Buttiker1,Camalet} and the non-equilibrium Green-Keldysh formalism. 
\cite{SL,Jauho,Guo,Arrachea} The scattering approach was primarily designed to describe the adiabatic pumping,   
in which the driving potential varies very slowly. The timescale on which the applied signal varies significantly
 exceeds the time needed for the electron to pass
through the system. The key result of the adiabatic scattering is a current formula in terms of an 
instantaneous (frozen) $S$ matrix. This matrix is computed perturbatively and only the linear term in
frequency is usually retained. A rigorous mathematical treatment of adiabatic quantum pumping \cite{Avron1} 
recovered
the BPT formula given in. \cite{Brower} Within this framework the relation between resonant transmission and
quantized pumped charge in an unbiased turnstile was analyzed.\cite{Aharony} An extended scattering formalism 
for studying both adiabatic and
 non-adiabatic quantum pumping was developed in Ref.\ \onlinecite{Buttiker1} and uses the Floquet theory and an 
$S$ matrix depending on two-energies. Recently Mahmoodian {\it et al.} \cite{Braginsky} have computed the
stationary current for a quantum wire submitted to alternating $\delta$-like voltages. It was shown there that
the current displays multiphoton peaks as a function of the Fermi momentum of the leads.     

Using the Green-Keldysh formalism Q-f Sun and T. S. Lin \cite{SL} have computed the current 
through a single level quantum dot when 
rectangular or harmonic potentials are applied at the contact to the leads. As it is well known, 
this model 
is exactly solvable within the wide-band limit (WBL) approximation  since the Dyson equation for 
the retarded Green
function is greatly simplified due to the leads' self-energy which within WBL is simply 
a delta-function. \cite{Jauho} Later on Wang {\it et al.} used the Keldysh approach to investigate the
non-adiabatic charge pumping in the presence of photon-assisted tunneling. 
The explicit calculation was done within WBL and for an unbiased double barrier pump 
driven by a local sinusoidal signal. It was shown that at large frequencies a nonvanishing current is 
generated even with a single-parameter perturbation. Moreover, a sign change of the pumped current
was reported when the Fermi level of the leads crosses the eigenvalues of the system. This feature
was predicted also by B\"{u}ttiker and Moskalets. \cite{Buttiker1}
 
Further progress was achieved by L. Arrachea for tight-binding models and periodic potentials. 
\cite{Arrachea} 
The method developed in this paper allows the calculation
of the d.c. component of the pumped current once the partial Fourier transform (i.e.\ the Fourier transform 
with respect to one time only) of the Green functions 
is known. It was shown that the d.c. component can be related to a transmission function $T(\omega)$ which is 
interpreted as the difference between the probabilities of tunneling out from and into the system. The numerical 
simulations are performed for unbiased one-dimensional pumps and the pumping potential is described by a diagonal  
time-dependent term added to the energy at the contact sites. For two harmonic potentials the retarded Green
function is computed perturbatively for weak pumping amplitude and pumping frequency.
The connection between the Floquet scattering and the Green-Keldysh function formalism for time-dependent transport
was discussed in Ref.\onlinecite{AM}. 

Besides the scattering theory and NEGF approach the pumping problem can be addresed via time integration 
of the Schr\"{o}dinger equation by the Crank-Nickolson 
approximation, as proposed by Stefanucci {\it et al.} in a series of papers.
\cite{Stefanucci,Stefanucci1} Their setup starts from the ground state of the unperturbed but 
{\it coupled} system and has therefore the advantage of introducing naturally the bias as a perturbation. 
We remind that the Keldysh formalism requires a partitioning of the system into `central region' 
and  `leads', the perturbation being the coupling between them which is established usually adiabatically 
in the remote past. \cite{Caroli}  
The Keldysh approach is however appropriate for studying the transport in the turnstile configuration
which requires to connect and disconnect periodically the pump from the leads. 

Recently the equation of motion method was applied to compute various
currents in 1D pumps coupled to finite wires with constant chemical potentials.\cite{AgSen} 
Such an assumption is
questionable for finite systems, especially when one is interested in the long time behavior.

In this work we are primarily interested in the transient effects on the transport properties of a 
many-level quantum dot turnstile which is submitted 
to pumping potentials at $t=0$. We believe these effects could seriously affect the transport properties
of nanostructures driven by fast oscillating signals.
Also, we mention that most of the previous theoretical approaches present calculations of the {\it stationary}
averaged current which implies either to look at the long-time limit behavior \cite{Stefanucci}, either to consider 
small pumping frequencies.\cite{Arrachea} Since in the stationary regime the transient effects are presumably 
washed out, one does not need
to specify
how and when the pumping signal is turned on. Note also that for a proper application of the Keldysh formalism it 
is crucial to have a well-defined equilibrium state of the decoupled system. 

While in the adiabatic regime the main advantage is to express transport
quantities in terms of the frozen scattering matrix up to errors of ${\cal O}(\omega^2)$ or even  
${\cal O}(\omega^3)$ (see the higher order corrections in Ref. \onlinecite{Levinson}), the NEGF formalism 
covers the 
entire frequency range, allowing therefore an equal footing treatment of adiabatic and nonadiabatic
pumping. The Green functions needed in the current formula are computed using a recently 
developed method \cite{us} which solves the integral Dyson equation {\it exactly} by transforming
it into an algebraic equation. Through this procedure the Green functions are computed taking into account
all back-and-forth scattering processes.    

The content of the paper is divided as follows. Section II describes the model and gives the relevant equations,
as well as the considered pumping potentials; more details about the formalism should be traced back from
Ref. \onlinecite{us}. Section III is the main part of the paper and presents the numerical results and
their discussion. Conclusions are summarised in Section IV.

\section{The model}

Within the tight-binding model which is adopted throughout this work the Hamiltonian of the system contains
three terms: the semiinfinite leads ($H_L$), the quantum dot turnstile ($H_S$) and the 
time-dependent pumping signals $H_T(t)$: 
\begin{equation}
H(t)=H_S+H_L+H_T(t).
\end{equation}
$H_S$ has a usual tight-binding form
\begin{equation}
H_S=\sum_{m=1}^{N}(\epsilon_m+V_g)d_m^{\dagger}d_m+\sum_{\langle m,n\rangle }t_{mn}d_m^{\dagger}d_n.
\end{equation}
Here $t_{mn}$ are hopping terms,
$\langle m,n\rangle $ denotes nearest-neighbor summation over the
system sites. $\epsilon_m$ is the on-site energy and the diagonal
term $V_g$ simulates a plunger gate potential applied on the system.
$N$ is the number of sites in the dot.

Following the experimental setup from Ref. \onlinecite{TSP} we describe the oscillating tunneling 
barriers between the dot and the leads by time-dependent hopping terms ($l,r$ denote the left and the right lead):   
\begin{equation}
H_T(t)=\sum_{\alpha=l,r}
V_{\alpha}(t) (c_{i_{\alpha}}^{\dagger}d_{m_{\alpha}}+h.c).
\end{equation}
Here $c_{i_{\alpha}}$ and $c_{i_{\alpha}}^{\dagger}$ denote the annihilation/creation
operators on the $i$-th site  of the lead $\alpha$ which is connected via the nearest neighbor hopping 
$V_{\alpha}$ to the site $m_{\alpha}$ of the dot. Similarly, $d_{m_{\alpha}}$ and $d_{m_{\alpha}}^{\dagger}$ 
correspond to the site $m_{\alpha}$ of the dot which is coupled to the lead $\alpha$.

\begin{figure}[tbhp!]
\includegraphics[width=0.45\textwidth]{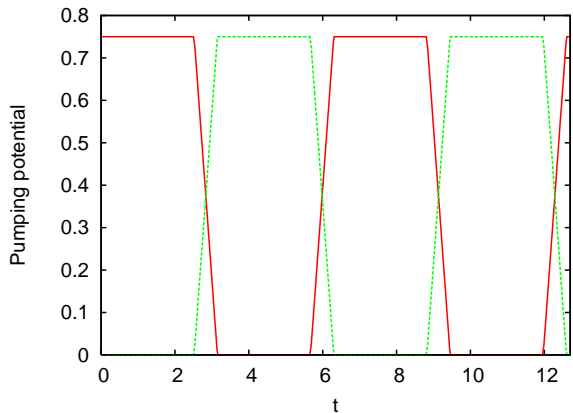}
\caption{(Color online) The pumping potentials applied on the left ($V_l$ - solid line) and right ($V_r$ - dashed line)
contacts.
We show two pumping cycles $k=2$. The parameter $\Delta=0.6$, $v_l=v_r=0.75$ and the frequency $\omega=1$.}
\label{figure1}
\end{figure}

The signals applied at the contacts between the dot and the leads have a trapezoidal form and are 
defined with the help of a function with period $2\pi$ introduced as follows ($\Delta$ is a positive number):
\begin{eqnarray}\label{cor1}
f(\omega t)=\left\lbrace \begin{array}{ccccc}
1   \qquad\qquad\qquad\qquad\qquad\,\,\,\, {\rm if}\,\, \omega t\in [0,\pi -\Delta], \\
1-\frac{1}{\Delta}(\omega t-(\pi-\Delta)) \qquad {\rm if}\,\, \omega t\in [\pi-\Delta,\pi],\\
0   \qquad\qquad\qquad\qquad\qquad {\rm if}\,\, \omega t\in [\pi,2\pi-\Delta], \\ 
\frac{1}{\Delta}(\omega t-(\pi-\Delta)) \quad\quad\,\,\, {\rm if}\,\, \omega t\in [2\pi-\Delta,2\pi].
\end{array}\right.
\end{eqnarray}
Then the pumping potentials of period $T$ are given by the relations:
\begin{eqnarray}\label{pump1}
V_l(\omega t)&=&v_{l}f(\omega t)\\
V_r(\omega t)&=&v_{r}(1-f(\omega t)),
\end{eqnarray}
where $\omega=2\pi/T$ is the frequency and $v_{l,r}$ are the amplitudes of the pumping signals. 
It is useful to introduce the number of pumping cycles $k$ considered in the numerical simulation. 
We show for clarity in Fig.\,1 a train of two such pulses (i.e. $k=2$) that we use to simulate the
turnstile configuration. The quantum dot is coupled suddenly to the left lead at $t_0=0$ while the right 
contact is off. 
 In the range $[kT/2-\Delta,kT/2]$ the sample is simultaneously isolated from the left lead
and connected to the right lead. This switching is done linearly at a slope $1/\Delta$ (note that a larger
$\Delta$ implies a slower onset of the couplings). In the second halfperiod of each pumping cycle 
the right contact is open, allowing thus the charge pumping. The cycle is completed by lowering the left
tunneling barrier (i.e. increasing $V_l(t)$) and turning off the coupling to the right lead. 

As it is widely known, the standard application of the Keldysh formalism leads to the following 
formula for the current entering the system from the left lead (here we take for simplicity one dimensional leads; 
a many-channel formula and more details are to be found in Ref.\ \onlinecite{us}): 
\begin{widetext}
\begin{equation}\label{current}
J_l(t)=-\frac{2e}{h}{\rm Im} (
\int_{-2t_L}^{2t_L}dE \int_{0}^tdse^{-iE(s-t)}
\Gamma^{l}(E;t,s)
(G_{ll}^R(t,s)
f_l(E)+G_{ll}^<(t,s)) ). 
\end{equation}
\end{widetext}
In the above formula the retarded and lesser  Green functions are given as usual in terms of 
Heisenberg operators
$G^R_{ll}(t,t')=-i\theta(t-t')\langle \{c_{i_l}(t'),d_{m_l}^{\dagger}(t)\}\rangle$
and $G^<_{ll}(t,t')=i\langle c^{\dagger}_{i_l}(t')d_{m_l}(t)\rangle$.
$f_l(E)$ is the Fermi function of the left lead, $t_L$ is the hopping energy on leads and 
$\Gamma^{l}$ is the linewidth depending on energy and time: 
\begin{eqnarray}\label{Gamma}
\Gamma^{l}(E;t,s)=\rho(E)
V_{l}(t)V_{l}(s),
\end{eqnarray}
containing the pumping potentials at different times and 
 the density of states at the endpoint of the semiinfinite one-dimensional lead $\rho(E)$:
\begin{equation}
\rho(E)=\theta(2t_L-|E|) \frac{\sqrt{4t_L^2-E^2}}{2t_L^2}.
\end{equation}
The retarded and the lesser Green functions are computed from
 the Dyson and Keldysh equations:
\begin{widetext}
\begin{eqnarray}\label{Dyson}
G^R(t,t')&=&G^R_0(t,t')+\int_0^tdt_1G^R(t,t_1)\int_0^{t_1}dt_2
\Sigma^R(t_1,t_2)G^R_0(t_2,t')\\\label{Keldysh}
G^<(t,t')&=&\int_0^tdt_1G^R(t,t_1)\int_0^{t'}dt_2\Sigma^<(t_1,t_2)G^A(t_2,t'),
\end{eqnarray}
\end{widetext}
where $G^{R,A}_0(t,t')$ are the retarded and advanced Green
functions of the isolated dot and $\Sigma^{R,<}$ are the
retarded and lesser self-energies. It worths mentioning that both self-energies 
contain the known Green functions of the semiinfinite leads but they are also quadratic functions 
of the pumping potentials whose time variable is different (see Eq. \ref{Gamma})
. Therefore, their time-dependence is
much more complicated than in other approaches were the pumping signals are applied to the system 
or to the leads and are described by diagonal terms in the Hamiltonian. In particular,  
the algorithm taken in Ref.\ \onlinecite{Stefanucci1} would we difficult to use. 
The time-dependent occupation number can be computed from the lesser Green function of the dot:
\begin{equation}
N(t)={\rm Im}\sum_{m=1}^NG^<_{mm}(t,t)=\sum_{m=1}^NN_m(t).
\end{equation}
$N_m(t)$ are on-site occupation numbers and will be used in the next section to gain information about the
internal charge dynamics during the pumping cycle.
A similar formula can be written down for the current $J_r(t)$ flowing from the system towards the 
right lead.

 As we have said, taking explicitely into account the initial instant when the pumping signals
are applied leads to transient effects. \cite{us} One consequence is that the period-averaged currents depend on the
period index $k$. We introduce therefore a $k$-indexed period-average for the currents (the $k$-th period covers the
interval $[t_{k-1},t_k]$ and $t_0=0$):
\begin{equation}
{\overline J}_{\alpha,k}=\frac{1}{T}\int_{t_{k-1}}^{t_k}dtJ_{\alpha}(t),\qquad \alpha=l,r.
\end{equation}
Although the approach taken in this work does not include the electron-electron interaction 
it captures the 
basic known features of turnstile pumps: the quantized pumped charge in low frequency regime and the 
satellite peaks due to photon-assisted tunneling in high-frequency regime.
Moreover, most of the 
results are presented at rather strong coupling to the leads when the dot is fairly open and 
the Coulomb blockade effects are not important. In a recent work Splettstoesser {\it et al.}\, \cite{S} 
proposed a method for dealing with Coulomb interactions in adiabatic quantum pumps. This approach is
based on the quantum Master equation and uses a perturbative expansion in the tunnel coupling.  
In the case of the unbiased quantum dot turnstile these authors found that one has to consider the second
order term in the tunnel coupling to get a non vanishing pumping.

\section{Numerical simulations}

In this section we present the main numerical results and discuss the 
transport properties of the turnstile pump in different regimes.
The bias, the frequency, the energy, the hopping constants on the leads, the coupling strengths and
the gate potential will be expressed in
terms of the hopping energy of the central region $t_D$ which is chosen as energy unit.
\begin{figure}[tbhp!]
\includegraphics[width=0.45\textwidth]{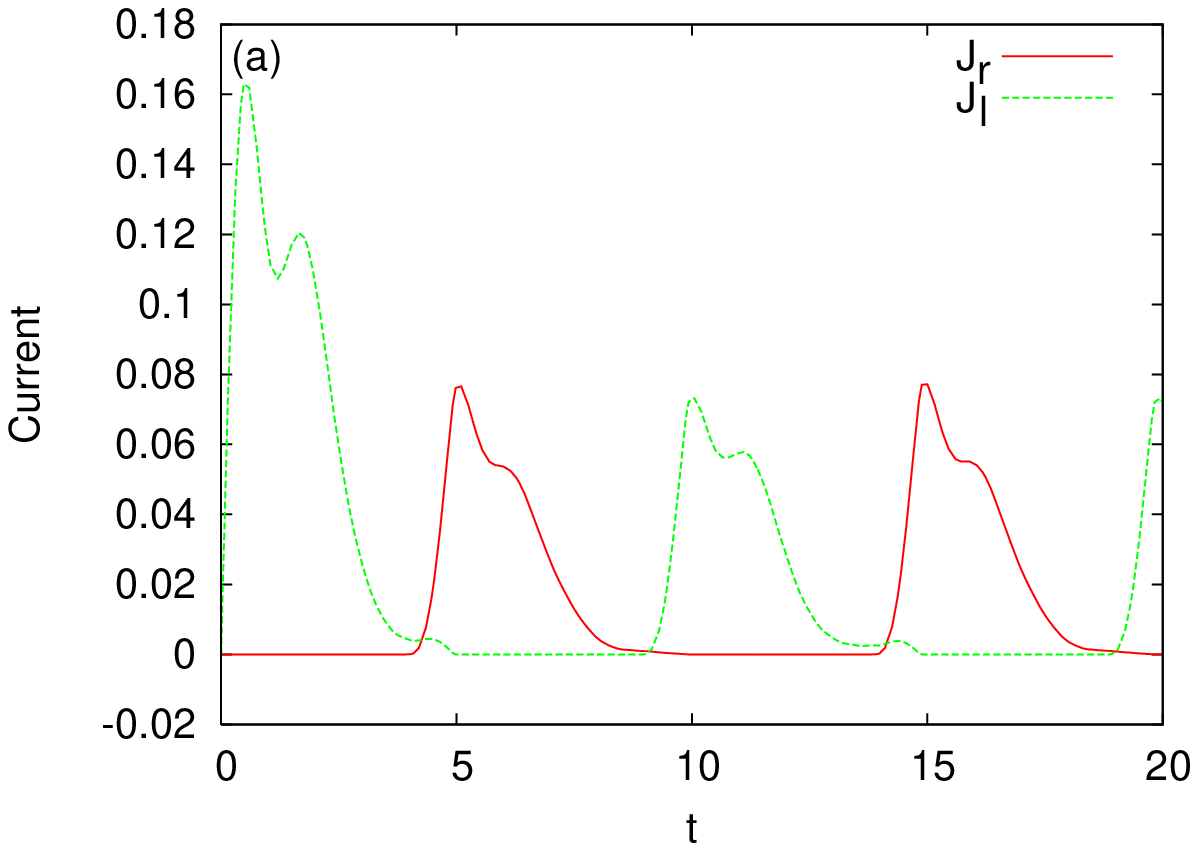}
\includegraphics[width=0.45\textwidth]{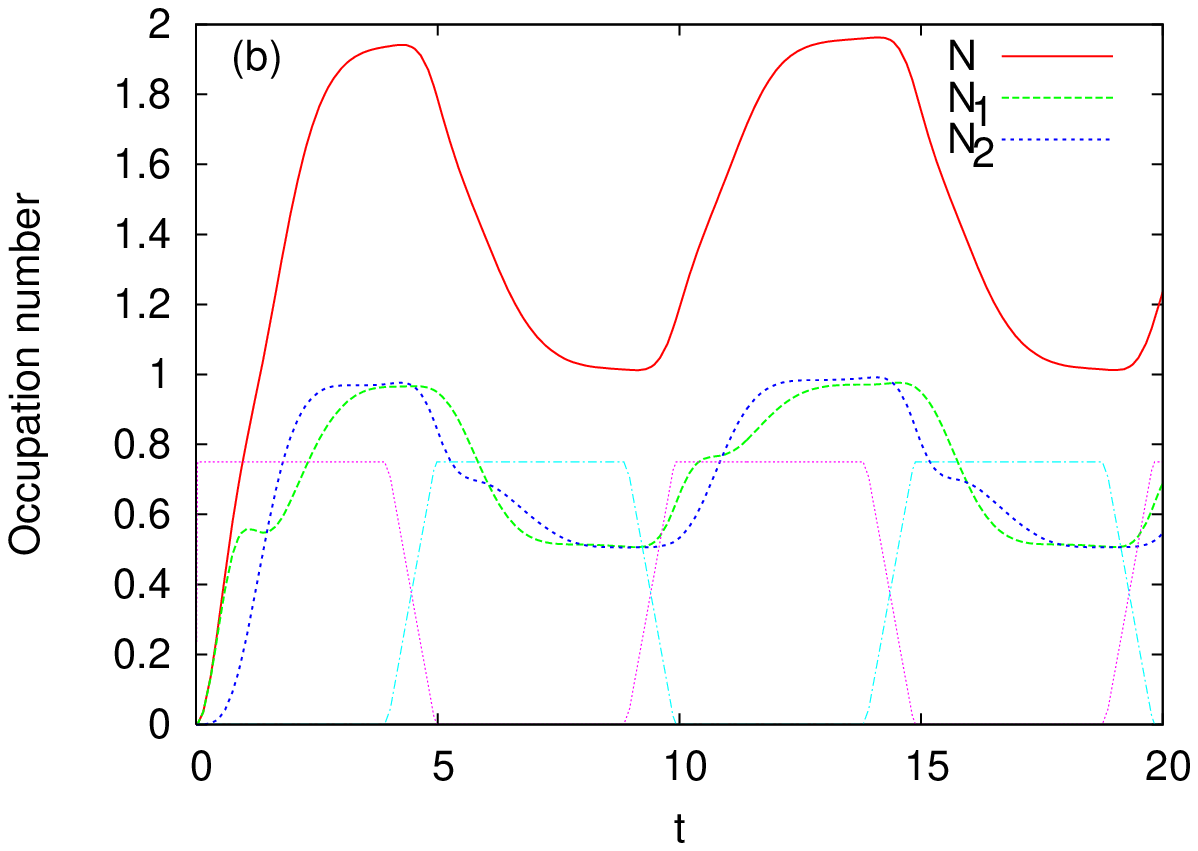}
\caption{(Color online) (a) The pumped current in the right lead ($J_r$) and the current entering the
system from
the left lead ($J_l$) for a 2-site turnstile.
(b) The occupation number of the dot $N(t)$ and the on-site occupations $N_i(t),\, i=1,2$. The
pumping potentials are also given.
During the second halfperiod of each pumping cycle the turnstile expells one electron
to the right lead. The bias is fixed to $W=3.0$, $\omega=0.6$, $v_l=v_r=0.75$ and $kT=0.0001$.}
\label{figure2}
\end{figure}

The current is therefore
given in units of $et_D/\hbar$ and the time expressed in units of $\hbar/t_D$.
 We take $e=\hbar=1$. The bias window (BW) is defined as the difference between the chemical potentials 
of the leads $W=\mu_l-\mu_r$. To make a connection to physical units one could 
take for example the energy unit as $t_D=0.1$ meV. Then the frequency unit would be $\omega\sim 25$ GHz 
and the time unit $t\sim 7$ ps. The termal energy $kT=0.0001$ in all numerical simulations.  

We consider first a two-site one-dimensional turnstile submitted to a finite bias and modulated by 
the trapezoidal signal introduced in Section II. In order to simulate the conditions of the
experiment performed by Kouwenhoven {\it et al.}\ \cite{TSP} we set the bias window to $W=3.0$ with 
respect to the chemical potential of the right lead $\mu_r=0$ such that the highest level 
of the isolated dot $E_1=1$ is located in it. Fig.\,2a shows the current $J_l(t)$ from the left lead 
towards the turnstile and the current $J_r(t)$ pumped into the right lead during two pumping cycles. 
The frequency is 
$\omega=0.6$ and the maximum height of the tunneling barriers is $v_l=v_r=0.75$. As expected, a nonvanishing 
current is generated in the right lead during the second halfperiod of each pumping cycle. The pumping
mechanism is proved by the behavior of the occupation number $N(t)$ which is given in Fig.\,2b.  
At the beginning of the first cycle the system collects charge from the left lead and since the 
coupling to the right lead is zero there are almost two electrons in the system after one halfperiod.
We shall call this halfperiod the charging halfperiod.
The charge dynamics within the system is also given in Fig.\,2b through the occupation numbers 
of the two sites $N_1$ and $N_2$. For the clarity of discussion we have also included in the figure the 
two potentials applied on the leads. 
The first site is rapidly populated up to 0.5 and then stays at this occupation while 
the second site starts to be filled. The current $J_l$ decreases in this short time range.

\begin{figure}[tbhp!]
\includegraphics[width=0.45\textwidth]{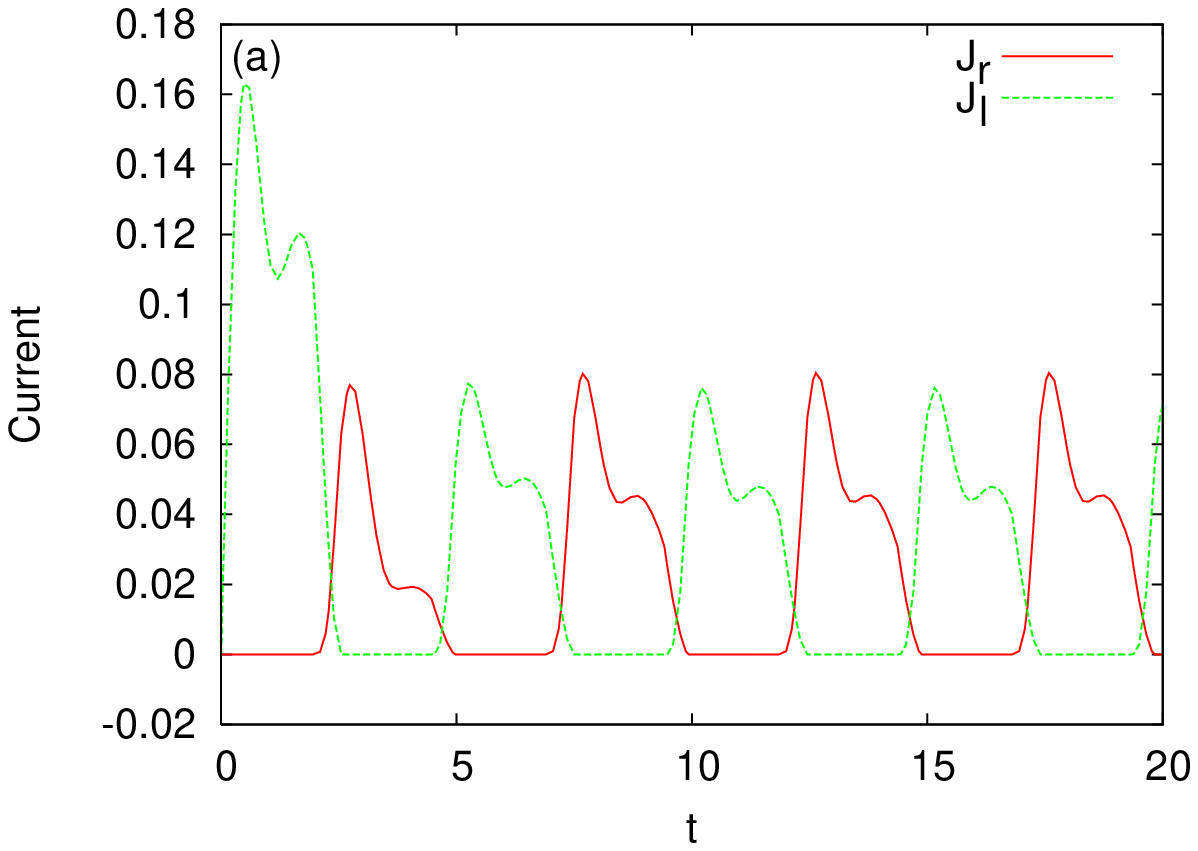}
\includegraphics[width=0.45\textwidth]{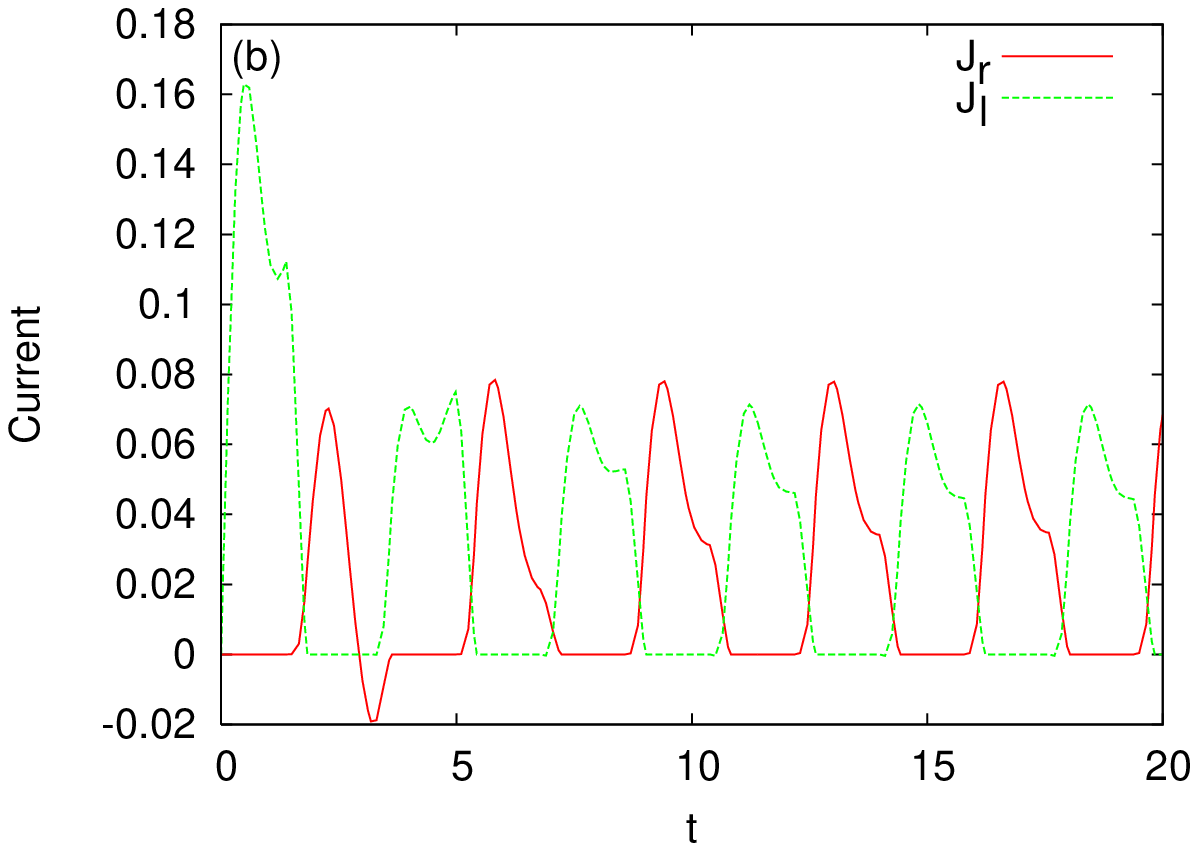}
\includegraphics[width=0.45\textwidth]{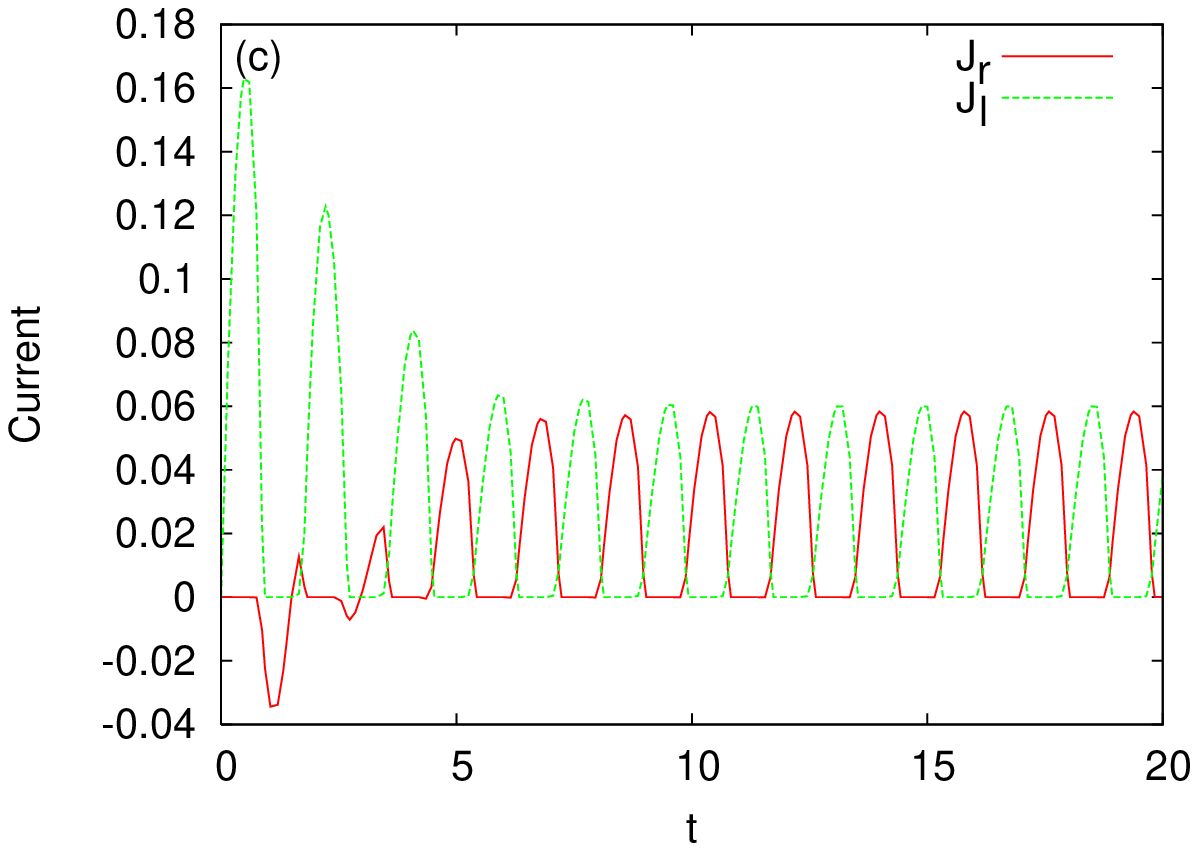}
\includegraphics[width=0.45\textwidth]{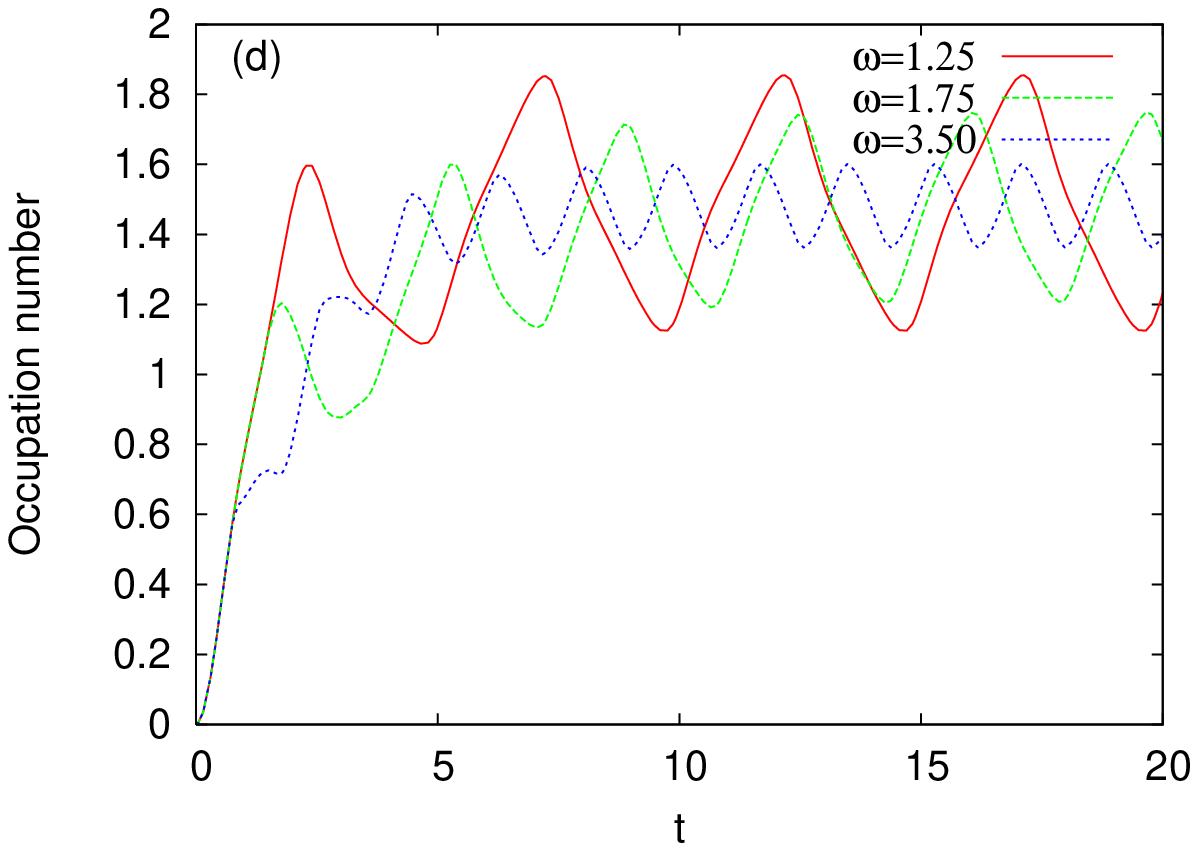}
\caption{(Color online) The shape of the currents $J_r$ and $J_l$ depends on the pumping frequency:
(a) $\omega=1.25$, (b) $\omega=1.75$, (c) $\omega=3.5$. At large frequency $J_r$ takes negative values
in the first pumping cycles. (d) The occupation number $N(t)$ corresponding to the three frequencies taken in
(a), (b), (c). $W=3.0$, $v_l=v_r=0.75$, $kT=0.0001$.}
\label{figure3}
\end{figure}

Then we see that $N_2$ increases faster than $N_1$ and both are reaching a constant value. Due to the
small frequency considered here the halfperiod of the pumping cycle exceeds the time needed for the 
system to be completely filled with electrons and therefore, as long as the coupling to the right lead is
still turned off, the total occupation number is steplike. The current $J_l$ becomes very small in this range
because as both sites are filled it is more difficult to inject charge (notice the lower slope at which
$N_1$ increases). The pumping starts effectively in the second half of the cycle
and leads to the transmission of one electron in the right lead, as observed in Ref. \onlinecite{TSP}. 
During the pumping halfperiod the occupation of the right contact drops quickly to 0.75 but then 
decreases at a lower rate as the first site is depopulating. Note that in the pumping sequence  
$N_1$ drops more rapidly than $N_2$.
The occupation number at the end of the pumping cycle does not decrease below 1 because the lowest level of the
dot $E_2=-1$ is well below the bias window and cannot contribute to transport.   

From the above discussion it is clear that the efficiency of the pump depends on two facts: first, the
charging halfperiod should allow the complete filling of the turnstile and second, the levels within the
bias window must be entirely depopulated during the pumping halfperiod. In the intermediate regime $\omega\sim V_{l,r}$
 (not shown) 
the system still transfers one electron, the difference being that the occupation
number decreases faster and more abruptly than in Fig.\,2. 
The transient effects on the 
time-dependent currents and pumped charge that appear as the pumping frequency increases are 
captured in Figs.\,3a-d. At $\omega=1.25$ the shape of the currents is similar to the one in Fig.\,2a.
However, the current pumped during the first cycle is smaller that the ones corresponding to the next
cycles and the occupation number plotted in Fig.\,3d shows that less charge is pumped. Also, 
there are no more steps in $N(t)$.    
The behavior of the occupation number helps us to identify the new features of the charge dynamics in 
the nonadiabatic regime. 

\begin{figure}[tbhp!]
\includegraphics[width=0.45\textwidth]{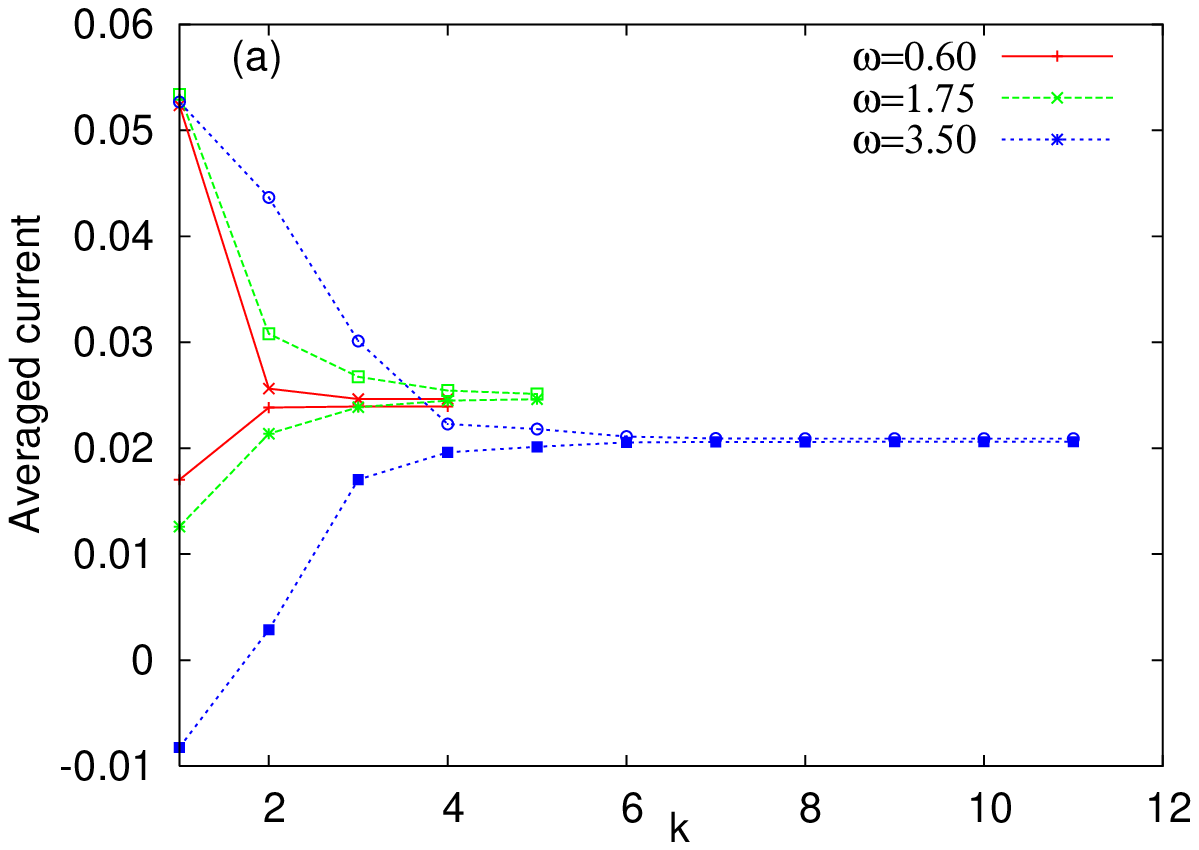}
\includegraphics[width=0.45\textwidth]{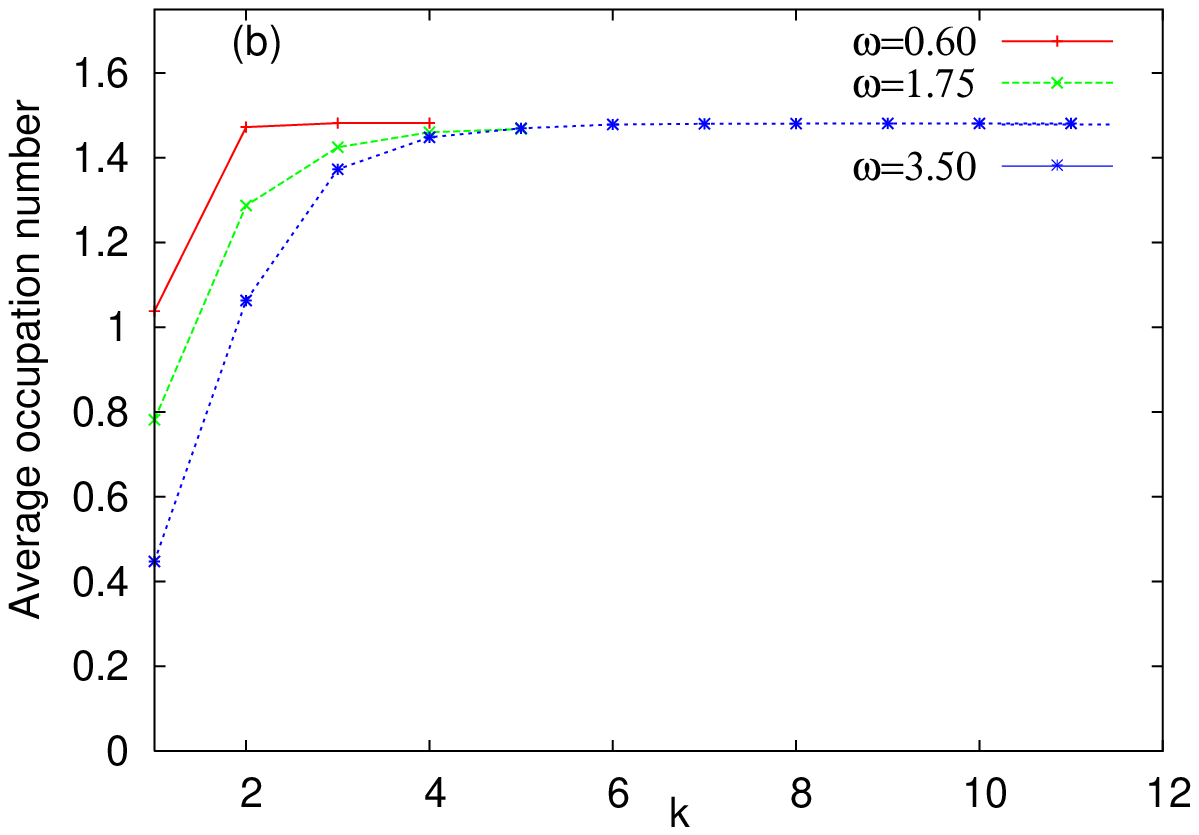}
\caption{(Color online) (a) The average currents ${\overline J}_{r,k}$ and ${\overline J}_{l,k}$ as a function of 
the period index $k$. There are three pairs of curves, corresponding to the frequencies $\omega=0.6$ ($k=4$), 
$\omega=1.75$ ($k=5$) and $\omega=3.5$ ($k=11$). For each frequency the two averaged currents are drawn 
with the same type of line (color); ${\overline J}_r$ (${\overline J}_l$) has a positive (negative) slope. 
(b) The average occupation number (see the discussion in the text).
$W=3.0$, $v_l=v_r=0.75$, $kT=0.0001$.} 
\label{figure4}
\end{figure}

On the other hand, the coupling to the right lead closes before
the level within the
bias window depopulates.
As a consequence, at the end of the pumping cycle there is a residual charge
($\sim$ 0.1) which is stuck within the bias window.
By further increasing the frequency up to $\omega=1.75$ another effect appears in Fig.\,3b.
Although there is a nonvanishing pumped current during the first cycle $J_r$ takes negative values at the
end of the cycle, which means that in this interval the system absorbs charge rather than pumps it.
Looking at the corresponding occupation number in Fig.\,3d one infers why the system does not act like a
pump over the entire pumping sequence. $N(t)$ goes slightly above 1 then drops to 0.9; during this interval the
system pushes electrons to the right lead and therefore $J_r(t)>0$. Then the occupation number increases,
leading to negative values of $J_r(t)$ (the pumping period in Fig.\,3b is 3.7).
The physical picture behind this is the following:
i) In the first
halfperiod both levels are populated, though not completely;
ii) During the first part of the pumping sequence
the level within the bias window depopulates, generating therefore a positive current $J_r$;

\begin{figure}[tbhp!]
\includegraphics[width=0.45\textwidth]{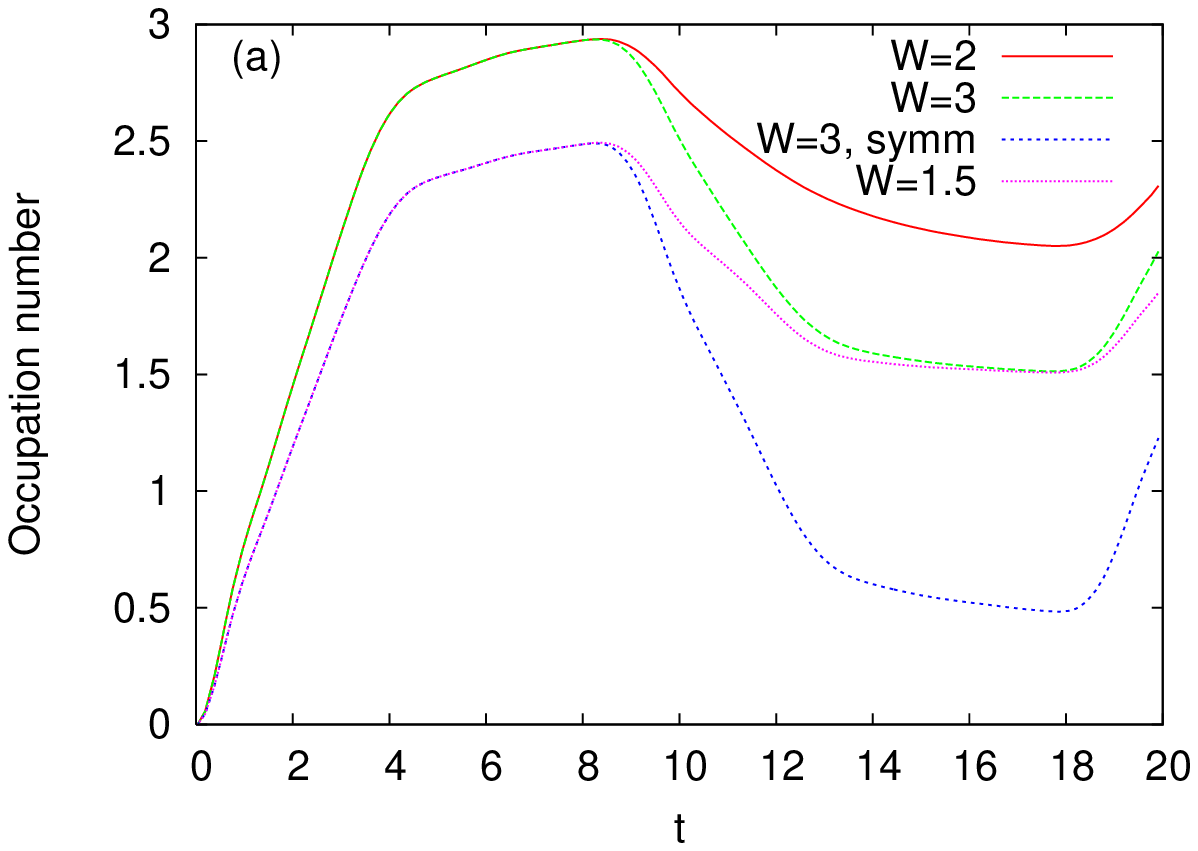}
\includegraphics[width=0.45\textwidth]{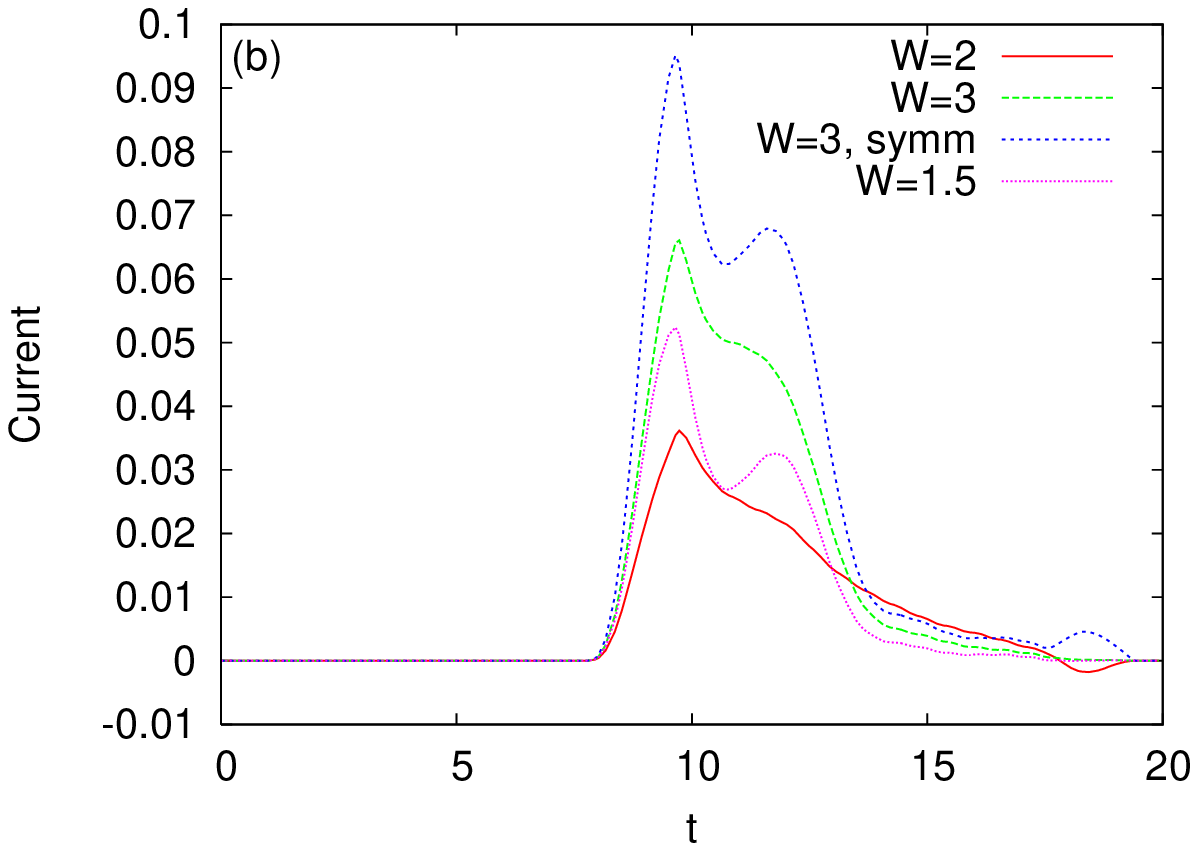}
\caption{(Color online) (a) The occupation number at  different values of the bias window and
different allignment of the turnstile levels with respect to the bias window.
(b). The pumped current $J_r(t)$ corresponding to the occupation numbers in (a).
 For all curves the frequency is set to $\omega=0.3$. We show only the first pumping cycle.}
\label{figure5}
\end{figure}

iii) When this
happens the total occupation number goes below 1; also, since the chemical potential of the right lead is higher
than the lower level of the dot, the system absorbs charge form the right lead and the occupation number increases
again. 

Finally, Fig.\,3c shows the currents in a highly non-adiabatic regime $\omega=3.5$. The steplike features
within each pumping cycle are washed out and on the first cycle the system does not pump any charge (the occupation
number is simply increasing and stays below 1). Moreover, even when a stable pumping regime is achieved
(for $t>5$) it is not effective at all since very little charge is pumped.

The transient regime should be noticed as well in the period-averaged currents. In order to check this
we give in Fig.\,4a these currents for the three frequencies considered in Fig.\,3.
The following things
are observed: i) In all three cases the d.c. components of the currents become eventually equal and
their value does not depend anymore on the period index. Moreover, the
averaged current is conserved. The passage to this `stationary' regime is faster
at low frequencies; ii) In the transient regime ${\overline J}_{l,k}$ exceeds ${\overline J}_{r,k}$ 
since there is
a net charge accumulation in the dot along each transient cycle (see the occupation numbers in Fig.\,3d); iii) At large
frequencies ${\overline J}_{r,k}$ takes negative values in the transient regime (see the first cycle at $\omega=3.5$) 
because the system absorbs
charge from the right lead.

Fig.\,4b shows that the average occupation number depends strongly on the period
index in the transient regime and settles down to 1.5 in the long-time limit. Notice that this means that 
 the level which contributes to the transport is half occupied.

The bias applied on the leads is an important parameter in turnstile operation because it 
controls the number of levels giving the main contribution to the current. As reported 
experimentally \cite{TSP} the number of electrons pumped during one pumping cycle is given 
by these levels. Fig.\,5a shows the occupation numbers of a three sites turnstile pump submitted to 
a different bias. 
We take a small frequency ($\omega=0.3$) and therefore the pumping cycle is long enough to allow 
the filling of the lowest levels (the dot has three eigenstates $E_{\pm}=\pm 1.4$ and $E_0=0$).  
For $\mu_l=3$ and $\mu_r=1$ only the highest level is located within the bias window and the pump transfers
one electron. The occupation number at the end of the cycle is $N=2$. We emphasize that the
frequency is much smaller than the gap between $E_0$ and $E_-$ and therefore the lowest level cannot 
give an important contribution to the current via excited sidebands.

By decreasing the chemical potential
of the right lead to $\mu_r=0$ the middle level alligns to $\mu_r$ and the turnstile pumps 1.5 electrons at $W=3$.
Note that due to the rather large coupling to the leads there is no charge quantization condition to be fullfilled.
 The dotted line shows the occupation number for the same bias $W$ with $\mu_l=1.5$ and 
$\mu_l=-1.5$. In this case we have a symmetric bias window (as marked in the figure) containing the middle 
level in the center.
The other two levels are close to
the chemical potentials of the leads. The transferred charge is $Q_p\sim 2$, suggesting that each level 
alligned to one chemical potential pumps only half electronic charge. This happens because the level close to
$\mu_r$ ($\mu_l$) are difficult to depopulate (populate). Also, by tuning the width and the position 
of the bias window it is possible to transfer the same charge in different ways (i.e.\, by involving different 
levels of the pump). For example, seting $\mu_l=1.5$ and $\mu_r=0$ one can still transfer one electron at a bias 
window $W=1.5$. Of course, the efficiency of the pump increases when the number of levels participating 
in transport increases.    
Fig.\,5b shows the pumped currents associated to the occupation numbers in Fig.\,5a.  
Although both cases $W=1.5$ and $W=2$ correspond to a pumped charge $Q_p=1$, the current is higher at 
$W=1.5$. Note also that the currents have a peak structure when two levels participate in transport.

The discussion around Figs.\,3 and 4 suggests that at a given frequency the transient regime would 
cover more pumping cycles if the number of levels located below the bias window increases. This is 
confirmed in Fig.\,6 which shows the occupation number of a 3-site turnstile. We take $\mu_l=3$ and 
$\mu_r=1$ and in this case the BW contains only the highest level $E_+=1.4$. 

\begin{figure}[tbhp!]
\includegraphics[width=0.45\textwidth]{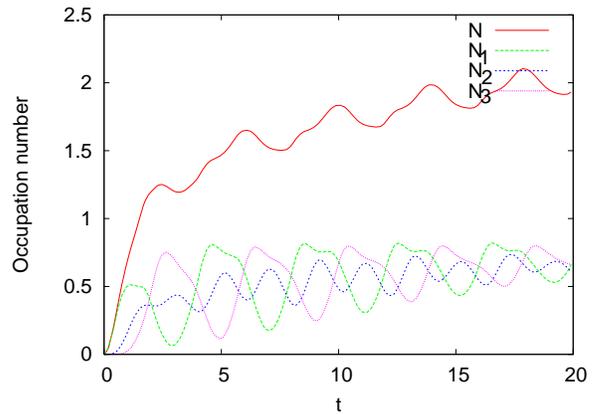}
\caption{(Color online) (a) The total occupation number $N(t)$ and the on-site occupation numbers $N_i$
($i,1,2,3$) for a 3-site turnstile submitted to
a bias $W=2$ and to a pumping signal of frequency $\omega=1.57$.}
\label{figure6}
\end{figure}

$N(t)$ shows clearly that
the other two levels are filled only after $k=5$ pumping cycles. Notably, even in the transient
regime the system pumps a small amount of charge ($\sim 0.25$) because of the highest level from the
BW. This level cannot be completely filled in the transient cycles because most of the
charge populates the lowest two levels. The behavior of the individual occupation numbers
$N_i$, $i=1,2,3$ confirms the intuitive picture of the internal charge dynamics. It is easily seen that
for any pumping cycle except the first, the occupation number of the 3rd site decreases even in the first
halfperiod, namely during the charging process. Since in this time range the contact to the right lead is
turned off the charge from the right contact can only flow back to the middle site. Indeed, $N_2$ increases
as $N_3$ decreases.

As the dot charges from the left lead the inverse flow is noticed. In the pumping halfperiod
the middle site occupation firstly increases because the charge accumulated in the first site passes to the right
contact which is now open, and finally decreases. All these internal bouncing trajectories are
taken accurately into account in our calculation because the method we used to solve for the Dyson equation
gives the {\it entire} matrix of the Green function (not only the diagonal matrix elements entering the current
formula).

\begin{figure}[tbhp!]
\includegraphics[width=0.45\textwidth]{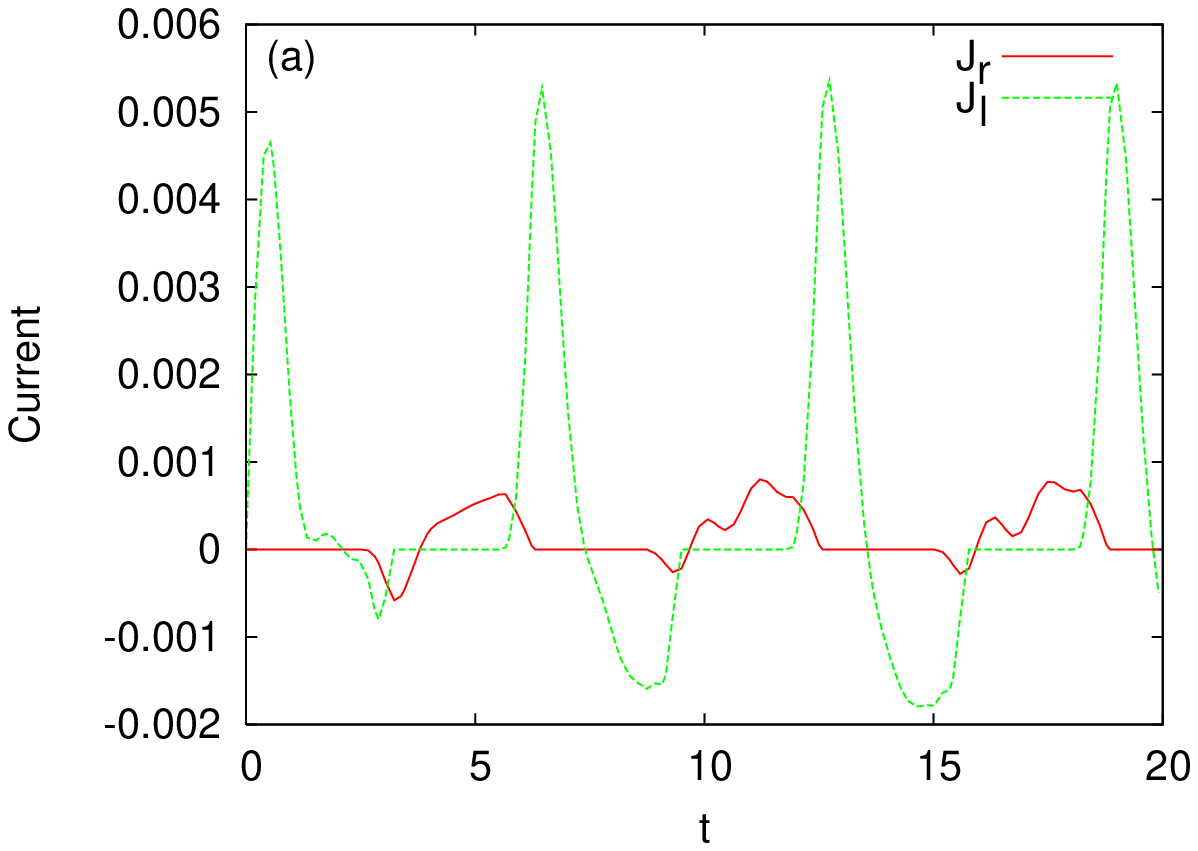}
\includegraphics[width=0.45\textwidth]{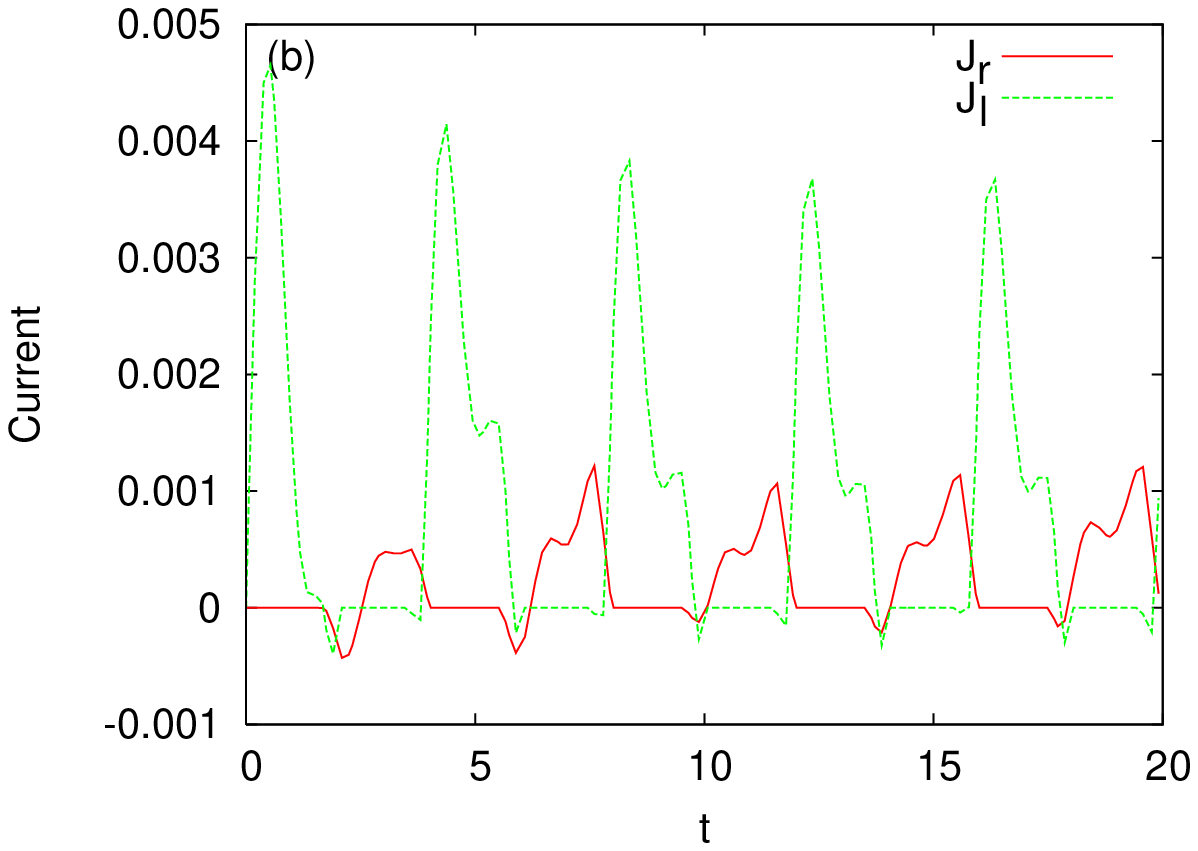}
\includegraphics[width=0.45\textwidth]{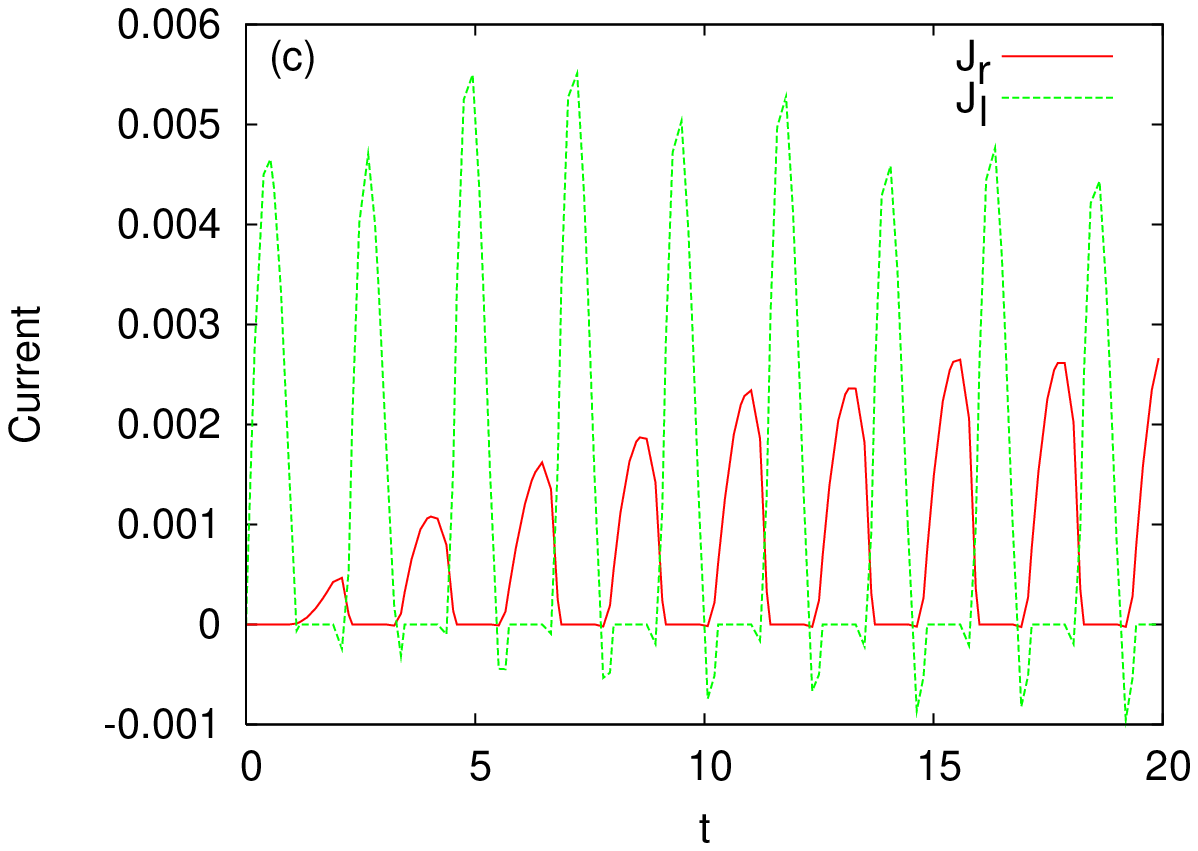}
\includegraphics[width=0.45\textwidth]{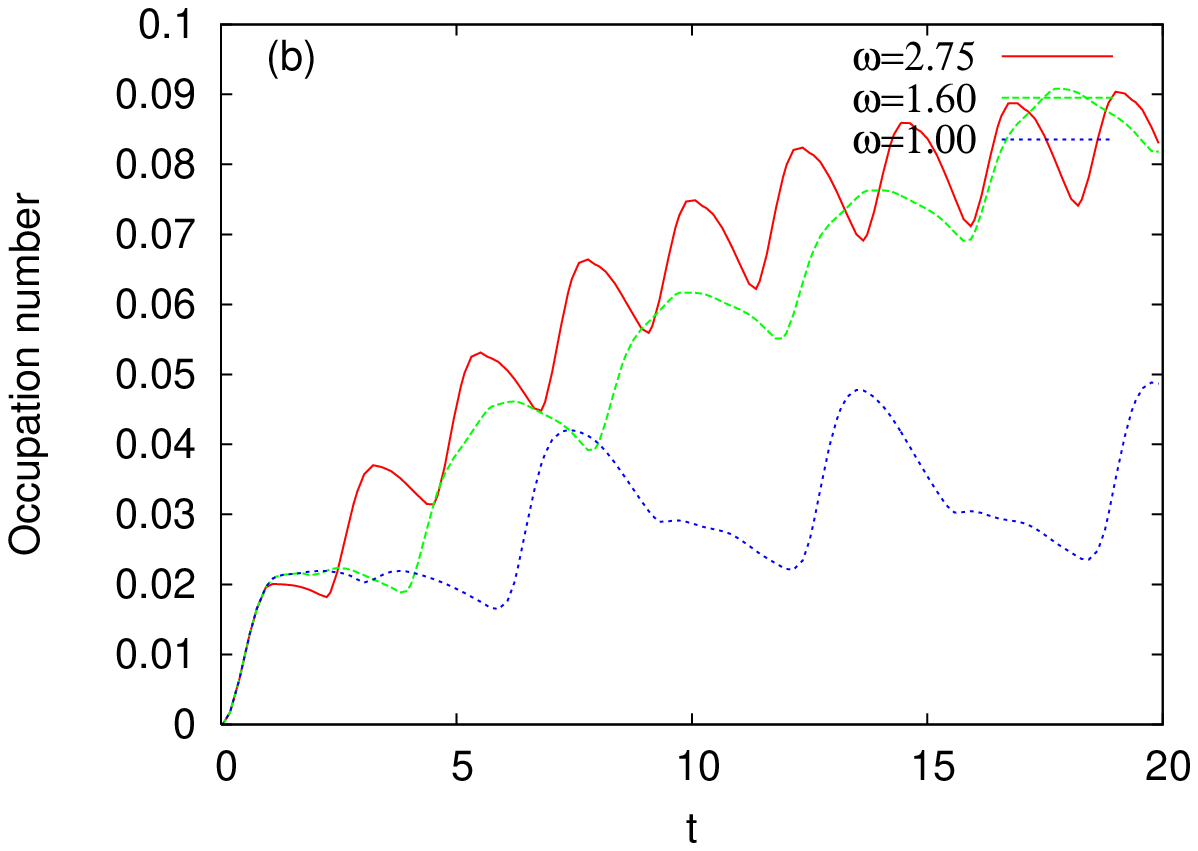}
\caption{(Color online) For a 3 site turnstile with all the levels above the bias window one can notice
 sideband contributions to the two time-dependent currents 
at frequency 
(a) $\omega=1.0$, (b) $\omega=1.6$ and
(c) $\omega=2.75$. (d) The occupation number at  different values of the frequency.}
\label{figure7}
\end{figure}

We turn now to another important feature of time-dependent transport, namely the contribution
due to inelastic scattering processes in which the incident electrons gain or loose energy quanta 
from the driving fields. Within the Floquet scattering approach it was shown \cite{Wagner,Buttiker1} that the  
outgoing electrons can have any energy $E_n=E+n\hbar\omega$ in the so called sideband ladder, $E$ being 
the energy of the
incident electrons. In the present model the pumping potential is {\it not} periodic because of the
condition $V_l(t)=V_r(t)=0$ for $t<0$, therefore the Floquet theorem cannot be rigorously applied.
At best, one can use results of the Floquet theory in the long time limit where all relevant quantities will 
oscillate with the frequency of the driving signals (see Ref.\, \onlinecite{Hone} for a discussion).
Nevertheless, inelastic tunneling processes are physically possible especially in the transient regime 
and should be noticed in the averaged currents. In particular one expects to see additional
contributions to the current at frequencies matching the gaps $\mu_{l,r}\pm E_i$ between the 
levels of the isolated system and the chemical potentials of the leads. 

In order to check these        
features we tune the bias window {\it below} the three levels of the 3-site pump by setting $\mu_l=-2.5$ 
and $\mu_r=-3.5$ such that $E_--\mu_l\sim 1$. We take a small pumping amplitude $v_l=v_r=0.35$ and 
look at the currents for several frequencies, as shown in Figs.\,7a-d. 
For small frequencies ($\omega=0.3$ and $\omega=0.5$ - not shown) there is no pumped current and $J_l$
takes both positive and negative values, while $N(t)$ increases and decreases accordingly during the charging
halperiod. This suggests that the system repels incident electrons back to the left lead. This happens
because on one hand there is no level below or within the bias window and, on the other hand, the frequency is 
too small to allow the population of the lowest level. The situation changes for $\omega=1$ (see  Fig.\,7a).
A pumped current appears in all three cycles presented. We related this to the occupation number plotted in
Fig.\,7d which shows that there is a charge pumped into the right lead. When comparing Fig.\,7a and Fig.\,7d
(the dotted line) one
notes that in the second half of the charging cycle $J_l<0$ and $N(t)$ decreases. This signals the 
energy relaxation process from the lowest level of the system to the left lead only, because the contact to 
te right lead is not yet open. Only the second (lower slope) decrease of $N(t)$ is associated to pumping trough 
the right lead. At $\omega=1.6$ (Fig.\,7b) the pumped charge and current increase. $J_l$ is mostly positive 
because the fast oscillating signals prevent relaxation to the left lead. The pumped current still shows a 
peak structure. In Fig.\,7c we show a highly non-adiabatic regime at $\omega=2.75$. In this case the second 
level $E_0=0$ will contribute as well. In order to avoid tunneling from the right lead to the level $E_-$ we have
considered for this curve $\mu_r=-5$. The peak structure of the pumped current dissapears and a periodic regime 
establishes slowly after 9 cycles. As we have said in the introduction, the present approach takes into
account all tunneling processes between the leads and the system and within the system. This nonperturbative
treatment might be crucial for describing the relaxation processes mentioned above, especially in the
transient regime where the resonances are not well defined.       

Finally we investigate briefly the satellite peaks appearing in the averaged current when a gate potential is
used to move the levels of the dot. These peaks were observed experimentally \cite{Kou} and are
associated with absorbtion and emission proces of energy quanta from the pumping fields. Theoretically
they were obtained first in the Master equation framework. \cite{Bruder} In the Keldysh formalism a calculation
of the stationary current in the WBL approximation was also presentent in Ref. \onlinecite{SWL}. 
Here we want to check wether the satellite peaks appear also in the first first pumping cycles, namely in the transient
regime.
To this end we consider the dependence of the pumped current for a two site turnstile on a gate potential 
$V_g$ (see the Hamiltonian 
given in the introduction) which shifts the two levels of the isolated system. Since we are interested in 
observing satellite peaks associated to one level only we take the parameters such that $v_{l,r}<
\omega<\delta2$, where $\delta=E_1-E_2=1$ is the level spacing. Fig.\,8a shows the averaged current associated to the 
2nd pumping cycle One notices as once two satellite peaks located on each side of main resonant peaks (at $V_g=\pm1$ 
the levels of the isolated dot are shifted to zero, threfore they are located in the middle of the bias window).  
Also, the distance between the sattelite peaks and the associated main peak equals roughly the frequency $\omega=0.65$
and confirms the absorbtion/emission picture. There are however several aspects due to the transient regime in which 
these peaks appear. i) As shown in Fig.\,8b the averaged current on the first pumping cycle  
does not display satellite peaks and is mostly negative. ii) Also, in Fig.\,8a the first sattelite peak has a 
negative value. This can be understood by looking at Fig\,8c which gives the 3D plot of the pumped current as
a function of time and gate potential. Clearly, around $V_g=-1.65$ there is a maximum positive current at the
beginning of the pumping cycle. In this range the system effectively pumps charge to the right lead via photon assited tunneling involving the highest level of the dot. This process is rather weak because in the transient regime the occupation
of the levels below the bias window is not complete. This is why after a short pumping regime the system absorbs
charge from the lead and threfore $J_r(t)<0$. As the first level enters the bias window a main peak appears
around $V_g=0$. Similar description can be made for the contribution of the second level. iii) For  
$V_g>0.3$ the average current is positive; the photon-assisted tunneling contribution to the 
pumping process is amplified and the current vanishes as the lowest level of the dot is pushed upwards. 

Our calculations show that photon assisted tunneling takes place also in the transient regime.     
 
\begin{figure}[tbhp!]
\includegraphics[width=0.45\textwidth]{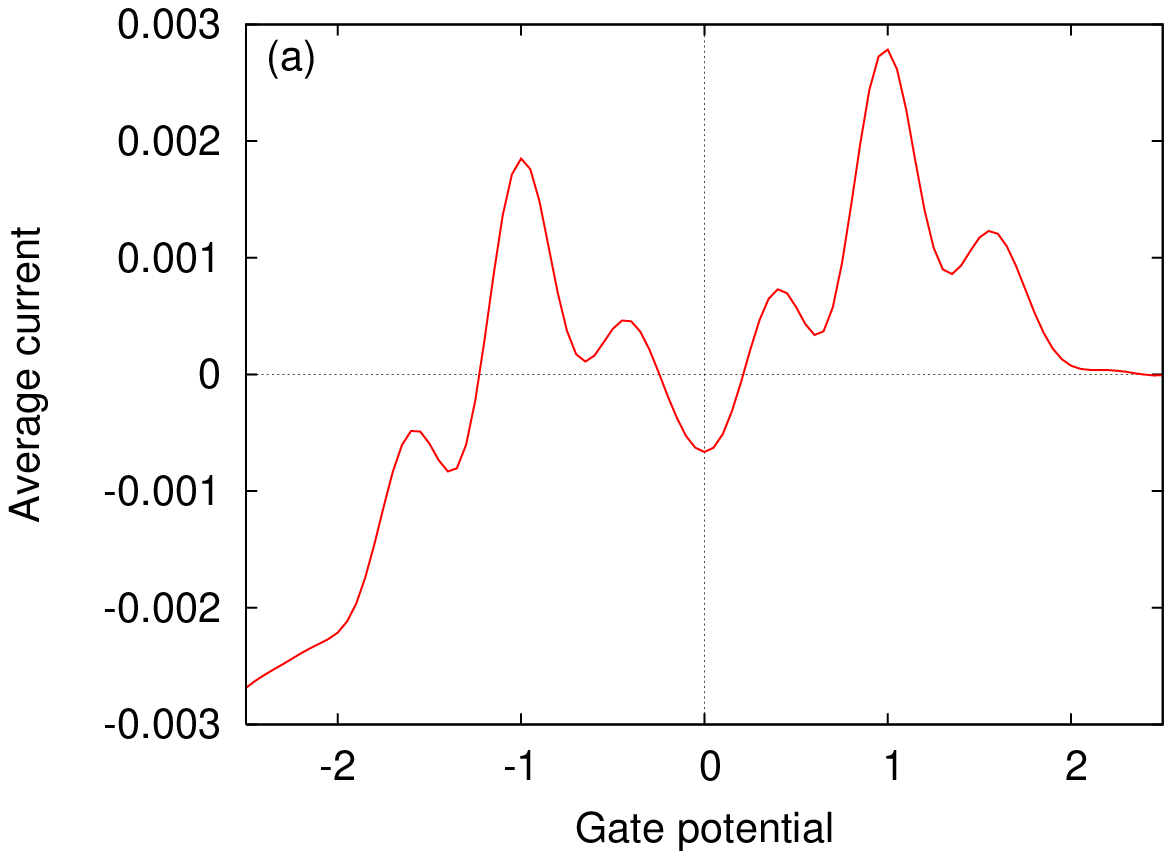}
\includegraphics[width=0.45\textwidth]{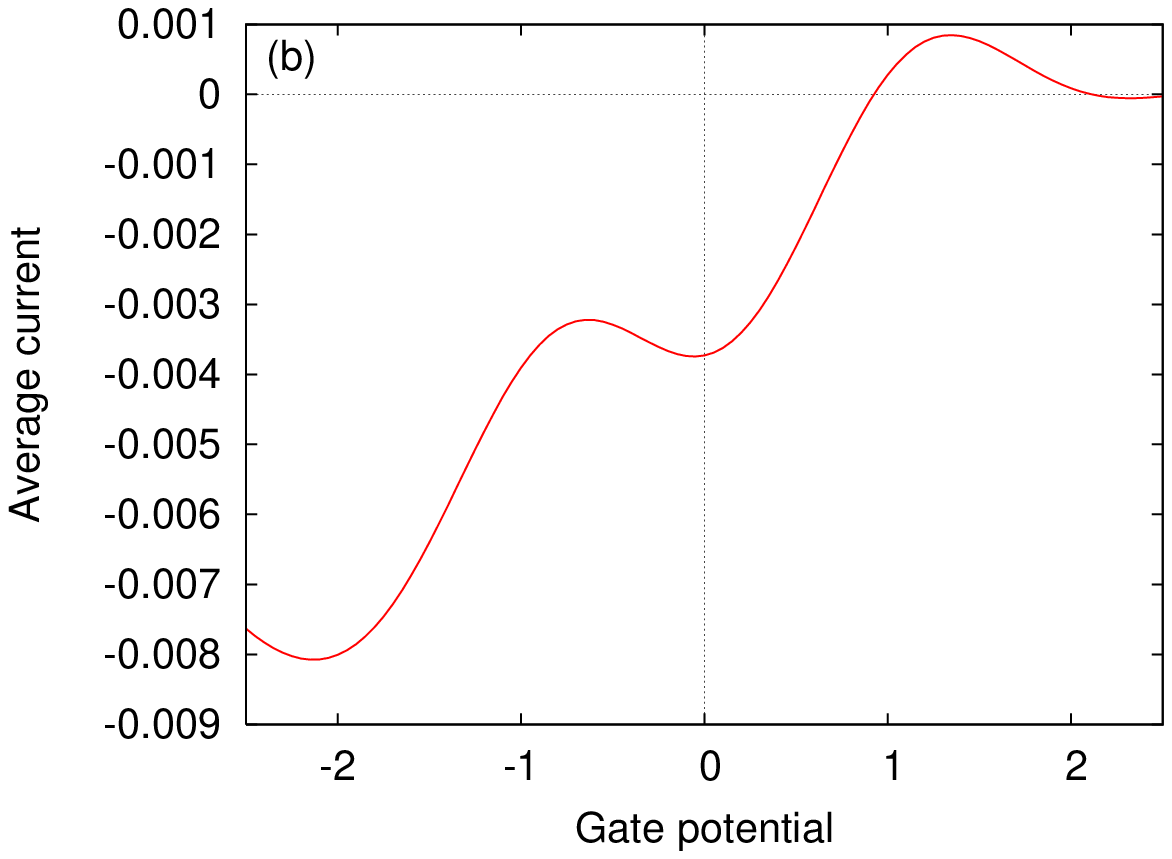}
\includegraphics[width=0.45\textwidth]{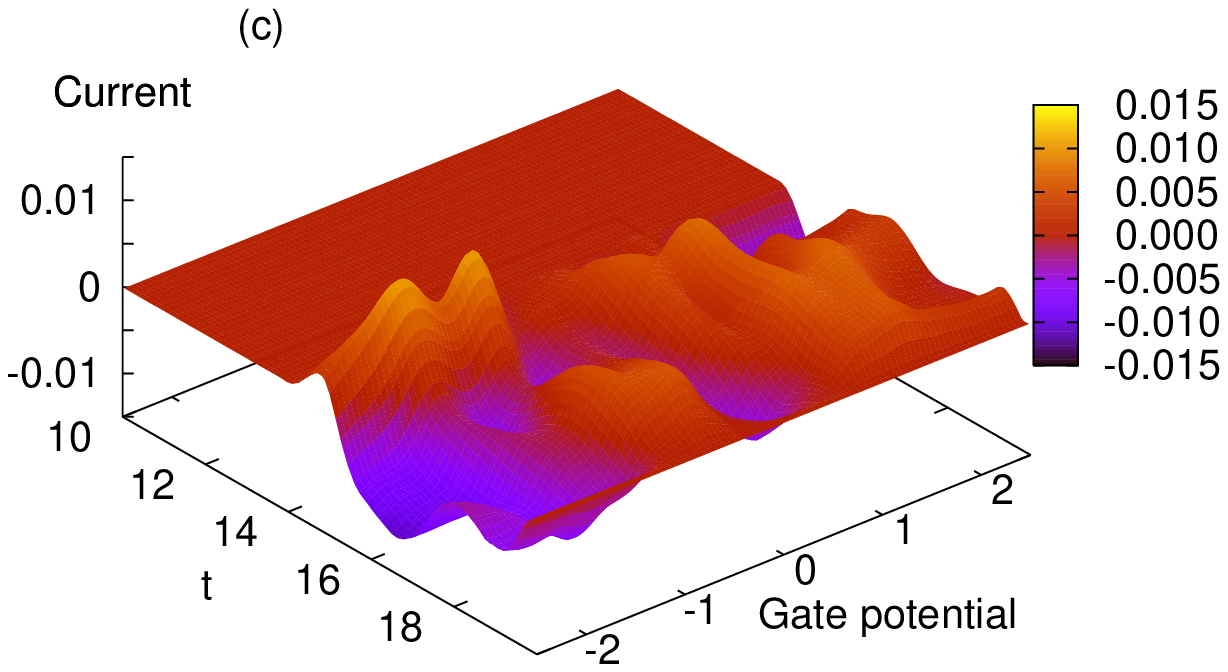}
\caption{(Color online) 
(a) The averaged current ${\overline J}_{r,2}$ as a function of the gate potential for a two
sites turnstile shows satellite peaks associated to photon-assisted tunneling during the second pumping cycle.
 (b) The averaged current corresponding to the first pumping cycle ${\overline J}_{r,1}$ does not contain
satellite peaks. 
(c) The current $J_r$ as a function of time (only the second cycle is shown) and gate potential. The discussion is
made in the text. Other parameters $\omega=0.65$, $v_l=v_r=0.35$, $\mu_l=0.2$, $\mu_r=-0.2$, $kT=0.0001$.}
\label{figure8}
\end{figure}

\section{Conclusions}

We have studied the turnstile pump regime of a few level noninteracting quantum dot within the
non-equilibrium Green-Keldysh formalism for time-dependent transport. 
The out-of-phase oscillating barriers coupling the system to the leads are described by time-dependent
hopping terms. In the numerical calculations we have considered a train of trapezoidal pulses that mimick
the configuration used in the experimental work of Kouwenhoven {\it et al.}\, \cite{TSP}.
 Our approach includes explicitely the starting time of the pumping cycles and therefore
captures the transient behavior of
the time-dependent and period-averaged current. To our best knowledge, such a calculation is presented here 
for the first time. We identify basically two stages of transport. The system experiences first a transient regime 
with poor pumping efficiency. This is due to the fact that the dot rather absorbs charge from the leads
in order to populate the levels located below the bias window. The number of pumping cycles in which the transient
features should be observed in future experiments with many level turnstiles increases if the bias window 
contains only the highest levels. In the second stage the occupation number
of the dot and the currents oscillate with the pumping period. 
 We show that  at low frequency and strong coupling to the leads an integer or half-integer
number of electrons are pumped, depending on the number and the location of the levels within the bias window.
In the high-frequency case the pumped charge is rather small even if additional contributions appear due
to scattering processes involving energy sidebands. We show that satellite peaks due to the photon-assisted     
tunneling appear also in the transient regime. The present analysis could be extended
to two-dimensional systems in order to discuss magnetic field effects. Also, different types of potentials
(e.g. harmonic or damped pulses) can be considered. 

\acknowledgments{This work was supported in part by the Icelandic Science and Technoloy Reasearch
Programme for Postgenomic Biomedicine, Nanoscience and Nanotechnology.
 V.M was also supported by CEEX Grant D11-45/2005. We acknowledge useful discussions with C. S. Tang.}


\begin{thebibliography}{18}

\bibitem{Switkes}
M. Switkes, C. M. Marcus, K. Campman, and A. C. Gossard, Science 283, 1905 (1999).
\bibitem{Geerligs}
L. J. Geerligs, V. F. Anderegg, P. A. M. Holweg, J. E. Mooij, H. Pothier, D. Esteve, 
C. Urbina and M. H. Devoret, Phys. Rev. Lett. {\bf 64}, 2691 (1990).
\bibitem{Kou}
W.G. van der Wiel, T.H. Oosterkamp, S. de Franceschi, C.J.P.M. Harmans, and L.P. Kouwenhoven.
Strongly Correlated Fermions and Bosons in Low-Dimensional Disordered Systems, eds. I.V. Lerner et al., pp.43-68
(2002), Kluwer Academic Publishers.
\bibitem{Zrenner}
H. J. Krenner, S. Stufler, M. Sabathil, E. C. Clark, P. Ester, M. Bichler, G. Abstreiter, J. Finley and A. Zrenner,
New J. Phys. {\bf 7}, 184 (2005).
\bibitem{Tarucha}
T. Fujisawa, D. G. Austing, Y. Tokura, Y. Hirayama, S. Tarucha,
J. Phys. Cond. Mat. {\bf 15}, R1395 (2003).
\bibitem{TSP}
L. P. Kouwenhoven, A. T. Johnson, N. C. van der Vaart, C. J. P. M. Harmans, and C. T. Foxon,
Phys. Rev. Lett. {\bf 67}, 1626 (1991).
\bibitem{Brower}
P. W. Brouwer, Phys. Rev. B {\bf 58}, R10135 (1998).
\bibitem{Levinson}
O. Entin-Wohlman, A. Aharony, and Y. Levinson, Phys. Rev. B {\bf 65}, 195411 (2002).
\bibitem{Buttiker1}
M. Moskalets and M. B\"{u}ttiker, Phys. Rev. B {\bf 66}, 205320 (2002)
\bibitem{Aharony}
O. Entin-Wohlman and A. Aharony, Phys. Rev. B {\bf 66}, 035329 (2002).
\bibitem{Camalet}
S. Kohler, J. Lehmann, P. Hänggi, Phys. Rep. {\bf 406}, 379 (2005).
\bibitem{Avron1}
J.E. Avron, A. Elgart, G.M. Graf, and L. Sadun, Comm. Pure and Appl. Mathematics LVII, 0538 (2004),
Phys. Rev. Lett. {\bf 87}, 236601 (2001).
\bibitem{Braginsky}
M. M. Mahmoodian, L. S. Braginsky, and M. V. Entin, Phys. Rev. B {\bf 74}, 125317 (2006), 
JETP {\bf 100}, 920 (2005).
\bibitem{SL}
Q-f Sun, T-h Lin, J. Phys. Cond. Matt {\bf 9} 3043 (1997).
\bibitem{Jauho}
A.-P. Jauho, N. S. Wingreen, and Y. Meir, Phys. Rev. B {\bf 50}, 5528 (1994).
\bibitem{Guo}
B. Wang, J. Wang, and H. Guo, Phys. Rev. B {\bf 68}, 155326 (2003).
\bibitem{Arrachea}
L. Arrachea, Phys. Rev. B {\bf 72}, 125349 (2005).
\bibitem{AM}
L. Arrachea, M. Moskalets, Phys. Rev. B 74, 245322 (2006).
\bibitem{Stefanucci}
  S. Kurth, G. Stefanucci, C.-O. Almbladh, A. Rubio, and E. K. U. Gross, Phys. Rev.
B {\bf 72}, 035308 (2005), G. Stefanucci, C.-O. Almbladh Phys. Rev. B {\bf 69}, 195318 (2004).
\bibitem{Stefanucci1}
G. Stefanucci, S. Kurth, A. Rubio, E. K. U. Gross, cond-mat/0701279.
\bibitem{AgSen}
A. Agarval, D. Sen, J. Phys. Condens. Matter {\bf 19} 046205, (2007).
\bibitem{Caroli}
C. Caroli, R. Combescot, P. Nozieres, and D. Saint-James, J. Phys. {\bf C} 4, 916 (1971).
\bibitem{us}
V. Moldoveanu, V. Gudmundsson, A. Manolescu, cond-mat/0703179.
\bibitem{S}
J. Splettstoesser, M. Governale, J. K\"{o}nig, and R. Fazio, Phys. Rev. B {\bf 74}, 085305 (2006).
\bibitem{Wagner}
 M. Wagner, Phys. Rev. A {\bf 51}, 798 (1995), Phys. Rev. {\bf B 49}, 16544 (1994).
\bibitem{Hone}
D. W. Hone, R. Ketzmerick, amd W. Kohn, Phys. Rev. B {\bf 56}, 4045 (1997).
\bibitem{Bruder}
C. Bruder and H. Schoeller, Phys. Rev. Lett. {\bf 72}, 1076 (1994). 
\bibitem{SWL}
Q-f Sun, J. Wang, and T-han Lin, Phys. Rev. B {\bf 58}, 13007 (1998).


\end{thebibliography}
\end{document}